\documentclass[%
reprint,
superscriptaddress,
 amsmath,amssymb,
prapplied,
]{revtex4-2}

\usepackage{graphicx}
\usepackage{dcolumn}
\usepackage{bm}
\usepackage{float}
\usepackage{color}
\usepackage[caption=false]{subfig}
\usepackage{siunitx}
\usepackage{tabularx}
\usepackage{array,multirow}
\usepackage[thinlines]{easytable}
\usepackage{physics}

\usepackage{hyperref}
\usepackage{xcolor}
\hypersetup{colorlinks=true, linkcolor=blue, citecolor=blue, urlcolor=blue}
\definecolor{DarkGreen}{RGB}{000,142,000}

\begin{document}

\title{Observation of symmetry breaking in time-varying scattering systems}

\author{M. Verde}
\email{email: miguel.verde@uam.es}
\affiliation{Departamento de Física Teórica de la Materia Condensada, Universidad Autónoma de Madrid, E-28049 Madrid, Spain}
\affiliation{Condensed Matter Physics Center (IFIMAC), Universidad Autónoma de Madrid, E-28049 Madrid, Spain}

\author{D. Globosits}
\affiliation{Institute for Theoretical Physics, Vienna University of Technology (TU Wien), 1040 Vienna, Austria}

\author{T. F. Allard}
\affiliation{Departamento de Física Teórica de la Materia Condensada, Universidad Autónoma de Madrid, E-28049 Madrid, Spain}
\affiliation{Condensed Matter Physics Center (IFIMAC), Universidad Autónoma de Madrid, E-28049 Madrid, Spain}

\author{C. M. Hooper}
\affiliation{Department of Physics and Astronomy, University of Exeter, Stocker Road, Exeter, Devon, UK, EX4 4QL}

\author{S. A. R. Horsley}
\affiliation{Department of Physics and Astronomy, University of Exeter, Stocker Road, Exeter, Devon, UK, EX4 4QL}

\author{S. Rotter}
\affiliation{Institute for Theoretical Physics, Vienna University of Technology (TU Wien), 1040 Vienna, Austria}

\author{P. A. Huidobro}
\affiliation{Departamento de Física Teórica de la Materia Condensada, Universidad Autónoma de Madrid, E-28049 Madrid, Spain}
\affiliation{Condensed Matter Physics Center (IFIMAC), Universidad Autónoma de Madrid, E-28049 Madrid, Spain}
\affiliation{Instituto Nicolás Cabrera (INC), Universidad Autónoma de Madrid, E-28049 Madrid, Spain}

\author{I. R. Hooper}
\email{email: i.r.hooper@exeter.ac.uk}
\affiliation{Department of Physics and Astronomy, University of Exeter, Stocker Road, Exeter, Devon, UK, EX4 4QL}

\begin{abstract}
From the Higgs mechanism to magnetic ordering, spontaneous symmetry breaking organizes physics across scales. Whether an analogous transition governs waves scattered by periodically driven structures has remained an open experimental challenge. Here we demonstrate a universal  symmetry-breaking transition in a periodically driven finite scattering system. By measuring the multispectral Floquet scattering matrix of a strongly modulated microwave resonator, we observe a transition from an unbroken symmetry regime with scattering eigenvalues of equal modulus to a broken symmetry regime where their moduli split apart, through an exceptional point. At parametric resonance, this transition culminates in coherent perfect absorption, where a tailored multispectral input state is completely absorbed without outgoing radiation, the singular limit of the underlying transition. Because it arises from the general algebraic structure of the Floquet scattering matrix rather than from implementation-specific details, this transition should occur broadly in such systems. Unlike instability phenomena in bulk driven media, it emerges in the eigenstructure of a finite scattering operator that couples waves across multiple frequency channels. Our results establish Floquet scattering as an experimentally accessible platform for the observation of symmetry breaking and for dynamic wave control in driven photonic systems.
\end{abstract}

\maketitle



\section{Introduction}

Symmetry breaking is one of the great unifying ideas of modern physics~\cite{Anderson1972,Beekman2019,Strocchi2008}.
It underpins phenomena as diverse as the Higgs mechanism in particle physics~\cite{Englert1964,Higgs1964}, magnetic ordering,
superconductivity, and crystallization~\cite{Chaikin1995}, as well as pattern formation in nonlinear systems~\cite{CrossHohenberg1993}. In wave physics, symmetry breaking has recently gained renewed prominence through exceptional points and parity–time-symmetric photonic structures, where transitions between distinct symmetry phases give rise to striking wave phenomena such as chiral modes, directional transport, and coherent perfect absorption~\cite{El-Ganainy2018,Ozdemir2019,Chong2010,Wan2011}.

In parallel, time-varying media have emerged as a versatile platform for controlling wave propagation \cite{Galiffi2022,Engheta2023,boltasseva2024photonic}. By exchanging energy with propagating fields, temporally modulated systems enable functionalities beyond the reach of static materials, including linear frequency conversion, parametric amplification, nonreciprocal transport, band-structure engineering and synthetic motion~\cite{Pendry2008,estep_magnetic-free_2014,yu-2009,bacot_time_2016,Li18,vezzoli2018optical,Li19,buddhiraju2020photonic,Koutserimpas20,Lyubarov2022,tirole,moussa2023observation,nasari_observation_2026}. Coherent wave control based on time interfaces as well as periodic temporal modulations has been demonstrated \cite{galiffi2023broadband,galiffi_optical_2026}. In bulk periodically driven media, such modulation can induce transitions between stable and unstable propagation regimes, caused by momentum gaps \cite{reyes2015observation,wang2025expanding}, and accompanied by amplification and lasing-like emission processes~\cite{wang2023metasurface,Kiorpelidis24,xiong2025observation,lee2026analogs}.

Spatially finite periodically driven systems, however, raise a fundamentally different question. In bounded scattering geometries, temporal modulation does not merely modify wave propagation inside the medium, but also couples incoming and outgoing fields across multiple frequency channels \cite{zurita2010resonances,Koutserimpas183,galiffi2020wood,Ptitcyn2022,CanosValero2026,garg2025photonic}. Their response is therefore governed by a multispectral Floquet scattering matrix, which captures the full scattering dynamics of the driven structure in frequency space~\cite{Li1999,Moskalets2002,Suwunnarat,Pantazopoulos2019,Buddhiraju2021,Fan2022,Globosits2024}.

Recent theoretical work revealed that this finite Floquet scattering problem supports a distinct and universal symmetry-breaking transition \cite{Globosits2026,garg2026BIC}. This universality arises because the transition follows from the general algebraic structure of the Floquet scattering matrix rather than from implementation-specific features such as geometry, dimensionality, or material realization. For periodically driven systems without loss or dispersion, the corresponding Floquet scattering matrix obeys a generalized conservation law associated with the number of pseudophotons~\cite{Pendry2023,Globosits2024} (or wave action~\cite{Brizard,Bellotti,Leonhardt}) rather than ordinary energy conservation. As a consequence, its eigenvalues can transition from an unbroken regime, where they remain unimodular and are thus located on the unit circle in the complex plane, into a broken regime, with eigenvalues leaving the unit circle. This transition is reached through exceptional points, where eigenvalues coalesce and the corresponding eigenvectors become parallel. At parametric resonance, the transition culminates in a singular limit in which one scattering eigenvalue vanishes while its partner diverges, corresponding to coherent perfect absorption (CPA) and its time-reversed counterpart, which is lasing.

These symmetry-breaking transitions in Floquet scattering systems are qualitatively distinct from instability phenomena in bulk driven media. They manifest through the eigenstructure of the multispectral Floquet scattering matrix, 
directly linked to experimentally measurable scattering channels, and naturally connect exceptional-point physics with CPA in a temporally driven setting. Whether this physics can be experimentally accessed remains an open challenge, however, since realistic resonant platforms inevitably involve dispersion, dissipation, finite bandwidth, and imperfect modulation, while the measurement of a multispectral scattering matrix poses substantial experimental challenges.

Here we experimentally demonstrate this universal Floquet scattering symmetry-breaking transition using a strongly modulated microwave resonator platform.
To this end, we introduce a theoretical framework for the Floquet scattering matrix of a dispersive, lossy time-varying scatterer, and establish the pseudounitarity of the Floquet scattering matrix as a fundamental property.
By reconstructing the full scattering matrix across coupled frequency channels, we directly track the movement of its eigenvalues as the modulation strength is increased and observe the transition from the unbroken to the broken symmetry regime through exceptional points. At parametric resonance, we additionally identify CPA as the experimentally accessible singular hallmark of this transition. Our results establish finite Floquet scattering as a practical platform for symmetry breaking and open new routes towards dynamic control of absorption, amplification, and wave interference in periodically driven photonic systems.

\section{Results and Discussion}

   \subsection{Symmetry breaking in time-varying resonators}
    
\begin{figure}[t!]
        \includegraphics[width=1.00\linewidth]{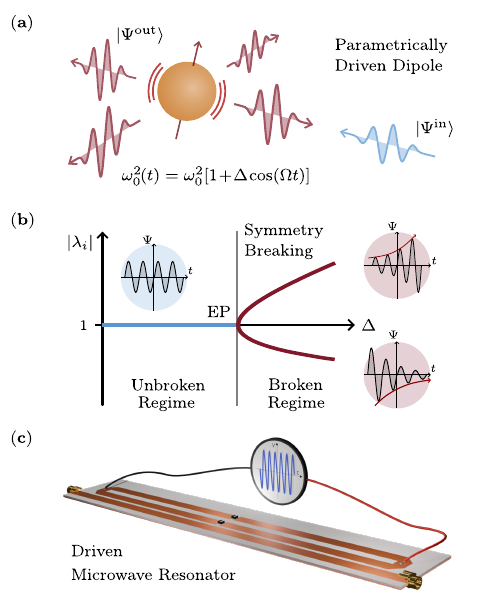}
        \caption{\textbf{Spontaneous symmetry breaking transition in time-varying media.}
        (a) Minimal model of a realistic time-varying medium: a driven and dispersive resonator, represented by an electromagnetic point dipole with parametric driving of its resonance frequency.
        (b) A periodic temporal modulation breaks energy conservation and the unitarity of the scattering matrix while giving rise to a new symmetry, \textit{pseudounitarity}, characterized by spontaneous symmetry-breaking transitions in its eigenvalues $\lambda_i$ and eigenvectors $\ket{\Psi}$, and inducing absorption and amplification. (c) Experimental platform for the observation of spontaneous symmetry breaking: a varactor-loaded ring resonator coupled to a transmission line.}
        \label{fig:1}
    \end{figure}

    Here, we demonstrate that spontaneous symmetry breaking transitions can arise in realistic time-varying open structures, which are finite and dispersive.
    To this end, we consider a simple model capturing the physics of a finite, lossy, dispersive, and time-periodic resonator: an electric point dipole whose resonance frequency is modulated periodically in time [see Fig.~\ref{fig:1}(a)].
    The dipole moment $\textbf{p}$ of such a generic scatterer satisfies the following damped, parametrically-driven differential equation
    \begin{equation}\label{eq:parametric}
        \dv[2]{\textbf{p}(t)}{t}+\gamma\dv{\textbf{p}(t)}{t}+\omega_0^2(t)\textbf{p}(t)= \kappa\textbf{E}(\textbf{x}_0,t)
    \end{equation}
    where $\textbf{E}$ is the incident electric field on the dipole, localized at $\textbf{x}_0$, and with coupling strength $\kappa$. 
    Furthermore, $\gamma$ is the intrinsic damping rate and $\omega_0^2(t)= \omega_0^2[1+\Delta\cos(\Omega t)]$ is the square of the time-periodic resonance frequency, with $\Omega$ and $\Delta$ the modulation angular frequency and strength, respectively, and $\omega_0$ the static resonance frequency.
    
    Given a time-dependent dipole embedded in free space, the electromagnetic field satisfies Maxwell's equations in vacuum, with a current density $\mathbf{J}$ localized at the position of the dipole.
    By writing Poynting's theorem for the parametric dipole introduced in Eq.~\eqref{eq:parametric}, and integrating over a spatial domain $\mathcal{D}$ that encloses the scatterer, we find the following energy balance relation,
    \begin{equation} 
        \dv{\mathcal{U}(t)}{t}+\int_\mathcal{D} \dd{\mathbf{x}} \nabla\cdot\mathbf{S}(t)
        =
        \frac{|\mathbf{p}(t)|^2}{2\kappa}  \dv{\omega_0^2(t)}{t} -\frac{\gamma}{\kappa}\left|\dv{\mathbf{p}(t)}{t}\right|^2.
    \end{equation}
    Here, $\mathcal{U}$ denotes the total energy stored in the modulated dipole and the electromagnetic field within $\mathcal{D}$, while $\mathbf{S}$ is the Poynting vector [see the Supplemental Material (SM)~\cite{SupplementalMaterial}].
    The first and second terms on the right-hand side denote the energy transfer induced by the temporal modulation and the energy dissipation through material losses, respectively.

    \begin{figure*}[t!]
        \includegraphics[width=1.00\textwidth]{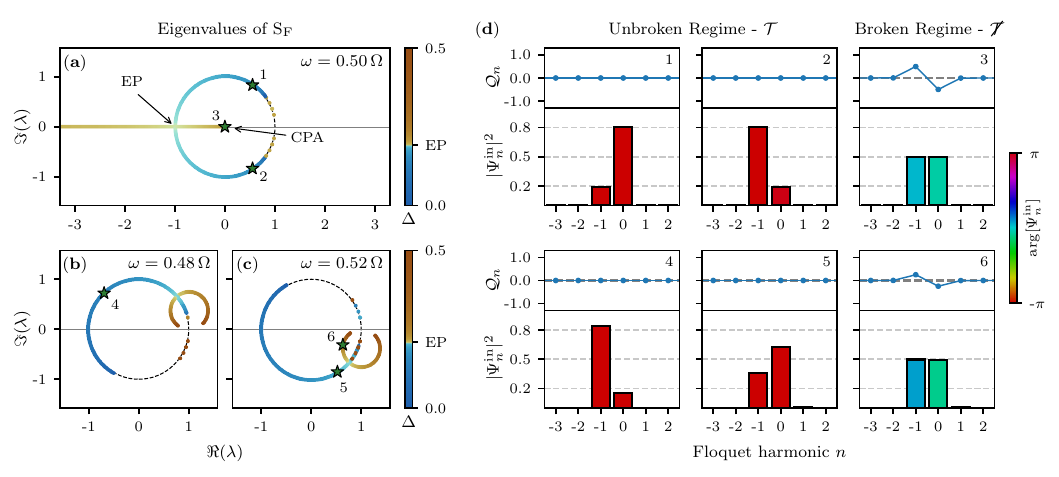}
        \caption{\textbf{Spontaneous symmetry breaking by a lossless parametric resonator.} (a-c) Properties of the eigenvalues $\lambda$ and (d) of the eigenvectors $\ket{\Psi^{\mathrm{in}}}$ of the Floquet scattering matrix $\mathrm{S}_\mathrm{F}$ of a dipolar resonator with a periodically time-dependent resonance frequency. The top and bottom rows correspond to the parametric resonance condition $(\omega=\Omega/2)$ and frequencies away from it $(\omega\neq\Omega/2)$, respectively. We track the eigenvalues of $\mathrm{S}_\mathrm{F}$ as a function of the modulation strength $\Delta$ for different input frequency $\omega=0.50\,\Omega$ (a), $0.48\,\Omega$ (b), and $0.52\,\Omega$ (c).
        For a weak modulation the eigenvalues remain on the unit circle (blue/beige colours) indicating that the system is in the unbroken regime.
        As the modulation reaches a critical value, EPs emerge where pairs of eigenvalues coalesce.
        Upon further increasing $\Delta$, these eigenvalues split and move away from the unit circle as inverse conjugate pairs. The associated eigenstates are now in the spontaneously broken regime (beige/red colours). Panel (d) presents the amplitudes (bars) and phases (colors) of the frequency components of the eigenvectors corresponding to six selected eigenvalues, marked by stars
        in panels (a)-(c).  In the unbroken regime (1,2 and 4,5), the eigenvectors are $\mathcal{T}$ invariant, with ${\Psi^\mathrm{in}_n}^*=\Psi^\mathrm{in}_n$, while in the broken regime (3,6), they are not.  
        The frequency components of the pseudophoton flux difference $Q_n$ [see Eq.~\eqref{eq:difference}] are also shown as dots joined by a guide-to-the-eye line. While all the components are zero in the unbroken regime, they may be non-zero in the broken regime, but pseudounitarity \eqref{eq:pseudounitarity} implies $\sum_n Q_n$ = 0 for all cases, such that the pseudophoton flux is always conserved. In the entire figure, the modulation frequency $\Omega=1.90 \,\omega_0$, the radiative damping rate $\gamma_\mathrm{rad}=\kappa\omega_0^2/(6\pi\varepsilon_0c^3)=0.037\,\omega_0$, and the material damping rate $\gamma=0$.}
        \label{fig:2}
    \end{figure*} 

   By Floquet's theorem \cite{Floquet1888}, the solutions to Eq.~\eqref{eq:parametric} may be written as a Fourier expansion over the frequency harmonics $\omega_n=\omega+n\Omega$, with $\omega$ the Floquet quasifrequency and $n\in\mathbb{Z}$. Accordingly, we expand any time-periodic quantity as
    \begin{equation} \label{eq:Floquet}
        \mathbf{F}(t)=\sum_n\mathbf{F}_n(\omega)\exp[-i\omega_nt].
    \end{equation}
   Using this representation and averaging Poynting's theorem over one modulation period, we obtain
    \begin{multline}\label{eq:wave-action}
        \int_\mathcal{D} \dd{\mathbf{x}} \nabla \cdot\sum_n \textbf{S}_n(\omega) = -\frac{\gamma}{\kappa}\sum_n\omega_n^2|\mathbf{p}_n(\omega)|^2
        \\ + \frac{1}{\kappa} \sum_{m,n}\omega_n\Im \left[  \mathrm{W}_{m}\,\textbf{p}_n(\omega)\cdot\textbf{p}^*_{n-m}(\omega)\right],
    \end{multline}
    where $\mathrm{W}_{m}$ represent the Fourier coefficients of the square of the (real) time-periodic resonance frequency, satisfying $\mathrm{W}_l=\mathrm{W}_{-l}^*$.
    Equation \eqref{eq:wave-action} makes it explicit that the net Poynting flux is not conserved due to the temporal modulation, even in the absence of material losses ($\gamma=0$).

    Thus, the natural question arises whether there exists a generalized flux that is conserved in a lossless dispersive time-varying system.
    Importantly, the balance law Eq.~\eqref{eq:wave-action} holds for each Floquet harmonic individually,
    which motivates the introduction of harmonics of a generalized flux $\mathbf{\tilde{S}}_n=\mathrm{K}_n\mathbf{S}_n$ whose weights $\mathrm{K}_n$ would cancel out the modulation-induced energy transfer.
    Interestingly, this condition is satisfied for $\mathrm{K}_n=\omega_n^{-1}$ (for details see SM \cite{SupplementalMaterial}). 
    This allows us to introduce the \textit{pseudophoton} (or \textit{wave action}) flux $\mathbf{\tilde{S}}(\omega) = \sum_n\omega_n^{-1}\,\Re[\textbf{E}_n(\omega)\times\textbf{H}_n^*(\omega)]$, 
    which, in the absence of loss ($\gamma=0$), satisfies 
    \begin{equation}\label{eq:wave-action-conserved}
        \int_\mathcal{D} \dd{\mathbf{x}} \nabla \cdot  \tilde{\textbf{S}}(\omega) = 0.
    \end{equation}
    While under a periodic temporal modulation the energy flux is no longer conserved, Eq.~\eqref{eq:wave-action-conserved} identifies a quantity that is conserved instead: the pseudophoton flux, corresponding to the conserved adiabatic invariant. 
     Crucially, here we show that the conservation of pseudophoton flux holds well beyond the non-dispersive systems for which it was originally identified~\cite{Pendry2023,Globosits2024}, and applies also to strongly dispersive resonators such as the parametric scatterer considered here.
     Thus, the pseudophoton flux is a conserved quantity of this lossless time-periodic medium, even if the system's response changes dramatically with frequency. 
    
    In a static system, conservation of energy flux is what renders the scattering matrix unitary~\cite{Newton2013}. For a time-periodic scatterer, the corresponding object is the Floquet scattering matrix $\mathrm{S}_\mathrm{F}$, relating the incoming and outgoing field amplitudes $\ket{\Psi^\mathrm{in}}$ and $\ket{\Psi^\mathrm{out}}$ in every harmonic frequency channel of the modulated resonator. Here, $\ket{\Psi^\mathrm{in,out}}$ are vectors containing the Floquet coefficients of incoming and outgoing electric fields, $\Psi^\mathrm{in,out}_n$.    

    The conservation of pseudophoton flux in our dispersive system then implies that $\mathrm{S}_\mathrm{F}$ is \textit{pseudounitary} also in this case (for details see the SM \cite{SupplementalMaterial}), satisfying
    \begin{equation}\label{eq:pseudounitarity}
        \mathrm{S}_\mathrm{F}^\dagger \mathcal{V} \mathrm{S}_\mathrm{F}=\mathcal{V}.
    \end{equation}
    Here, $\mathcal{V}$ is a signature matrix that assigns additional minus signs to the negative-frequency channels induced by the periodic modulation.
    Interestingly, such a pseudounitarity condition is reminiscent of parity-time ($\mathcal{P}\mathcal{T}$) symmetric systems \cite{christodoulides2018parity}, although we emphasize that the physical system under study here does not consist of spatially-balanced loss and gain elements \cite{Ruter2010}.

    \begin{figure*}[t!]
        \includegraphics[width=1.00\linewidth]{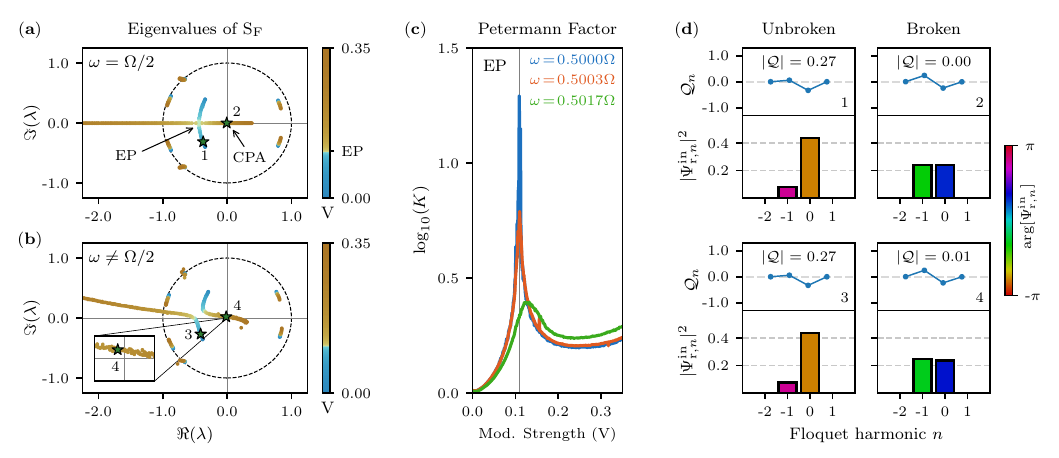}
        \caption{\textbf{Experimental observation of spontaneous symmetry breaking in a time varying microwave resonator.} Properties of (a)-(b) the eigenvalues $\lambda$ and (c)-(d) the right eigenvectors $\ket{\Psi_\mathrm{r}^\mathrm{in}}$ of the experimental Floquet scattering matrix $\mathrm{S}_\mathrm{F}$ of a varactor-loaded ring resonator with a time-periodic capacitance coupled to a finite transmission line. 
        (a)-(b) Evolution of the scattering matrix eigenvalues $\lambda$ as a function of the modulation voltage for input frequencies $\omega=0.5000\,\Omega$ (a), $0.5003\,\Omega$ (b). 
        While the intrinsic material damping of the experimental setup shifts the eigenvalues of the strongly interacting modes inside the unit circle for low modulation voltages, the overall behaviour with increasing modulation voltage is qualitatively similar to the lossless case: eigenvalues depart from an EP at the parametric resonance condition (a), or from an avoided crossing away from it (b).
        Note that the eigenvalues present close to unit circle correspond to frequency channels that do not take part in the parametric interaction mediated by the resonator.
        (c) Maximum Petermann factor $K$ of the scattering matrix, which diverges when two eigenvectors coalesce, is shown here as a function of the modulation voltage $(\mathrm{V})$ for different input frequencies $\omega=0.5000\,\Omega$, $0.5003\,\Omega$ and $0.5017\,\Omega$. 
        Only at parametric resonance, $\omega=\Omega/2$, does $K$ peak sharply (vertical line), signalling an EP; away from $\Omega/2$ the peak rapidly collapses. 
        (d) Signatures of spontaneous symmetry breaking in the Floquet components of the eigenvectors corresponding to the four starred eigenvalues in panels (a)-(b), along with the frequency components of the pseudophoton flux difference $\mathcal{Q}_n$.
        In the spontaneously broken regime (eigenvectors 2 and 4), the eigenvectors are clearly non-$\mathcal{T}$ invariant. 
        While material damping formally removes the pseudounitarity, implying that in general $\mathcal{Q}=\sum_n \mathcal{Q}_n \neq 0$, $\mathcal{Q}$ nonetheless remains $\approx$ 0 for the symmetry-broken eigenvectors 2 and 4, as a consequence of the underlying pseudounitarity, and in fact identically vanishes for eigenvector 2 at the parametric resonance condition (see discussion in the text).
        }
        \label{fig:experiment}
    \end{figure*}

    The pseudounitary condition \eqref{eq:pseudounitarity} of the Floquet scattering matrix implies that the system supports a distinct and universal symmetry-breaking transition~\cite{Globosits2026}.
    Indeed, while the eigenvalues $\lambda$ of a unitary matrix are unimodular, $|\lambda|=1$, pseudounitarity only enforces eigenvalues to come in inverse conjugate pairs $\{\lambda, 1/\lambda^*\}$ \cite{Mostafazadeh2001}.
    This allows the system to transition from an unbroken regime, where $\lambda$ is its own inverse conjugate and thus $|\lambda|=1$, to a broken regime, where $|\lambda|\neq1$.
    The onset of the broken regime reflects a spontaneous symmetry breaking transition occurring at an exceptional point (EP) where two eigenvalues coalesce and their corresponding eigenvectors become parallel.
    In particular, with the cosine modulation profile considered here, the time-reversal operator $\mathcal{T}$ maps eigenstates from the unbroken regime onto themselves, and eigenstates from the broken regime onto their partner state corresponding to the inverse conjugate eigenvalue.
    We note that the operator involved in the symmetry breaking is determined  by the modulation profile. The pseudounitarity of the Floquet scattering matrix and its consequences are, however, independent of the type of modulation, provided that it is periodic.

    Figure~\ref{fig:2} illustrates the implications of the conservation law Eq.~\eqref{eq:wave-action-conserved} and the associated symmetry-breaking transition through the properties of the Floquet scattering matrix. For the three-dimensional dipole scattering problem considered here, the Floquet scattering matrix reads $\mathrm{S}_\mathrm{F}(\omega)=\mathbb{I}+2\mathrm{G}_\mathrm{I}(\omega)\mathrm{A}(\omega)$, with $\mathrm{G}_\mathrm{I}$ being the matrix representation in Floquet space of the imaginary part of the free-space dyadic Green's function evaluated at the dipole's position, and $\mathrm{A}=(\mathrm{A}_\mathrm{B}^{-1}-\mathrm{G}_\mathrm{I})^{-1}$ being the dressed dipole polarizability that includes the dipole self-interaction, with $ \mathrm{A}_\mathrm{B}$ the bare polarizability stemming from Eq.~\eqref{eq:parametric} \cite{SupplementalMaterial}. In Fig.~\ref{fig:2}(a)-(c), we display the eigenvalues of $\mathrm{S}_\mathrm{F}$ in the complex plane as a function of the modulation strength $\Delta$, for $\omega=0.50\,\Omega,\;0.48\,\Omega,$ and $0.52\,\Omega$, respectively. The modulation frequency $\Omega=1.90\,\omega_0$ is fixed close to twice the resonance frequency. 
    In all three cases, a weak modulation strength keeps the eigenvalues unimodular as they remain on the unit circle, so that the system stays in the unbroken regime (blue/beige colors).
    As the modulation strength reaches a critical value (beige color), the system undergoes a spontaneous symmetry-breaking transition and EPs emerge. 
    By further increasing $\Delta$, the eigenvalues split into inverse conjugate pairs and move away from the unit circle, corresponding to the broken regime (beige/red colors).
    In particular, under the parametric resonance condition $\omega=\Omega/2$ (Fig.~\ref{fig:2}(a), top panel), one eigenvalue goes to zero while its inverse conjugate diverges, allowing the dispersive scatterer to simultaneously exhibit CPA and lasing. In the calculations, we use 10 Floquet modes: a pair of them is responsible for the symmetry breaking transition, while the others remain on the unit circle.

    This spontaneous symmetry breaking transition also manifests in the Floquet scattering matrix eigenvectors.
    The Floquet components of the eigenvectors, $\Psi^\mathrm{in}_n$, corresponding to the six eigenvalues marked with a star in Fig.~\ref{fig:2}(a)-(c) are shown in Fig.~\ref{fig:2}(d).
    Four of these eigenvectors (1,2,4,5) lie in the unbroken regime and are thus invariant under time-reversal symmetry, i.e., complex conjugation ${\Psi^\mathrm{in}_n}^*=\Psi^\mathrm{in}_n$.
    The remaining two eigenvectors (3,6) lie in the broken regime and are thus not invariant under this symmetry operation, a signature of the spontaneously broken symmetry.

   The conservation law Eq.~\eqref{eq:wave-action-conserved} tells us that the pseudophoton flux flowing into the system must equal the pseudophoton flux flowing out in both the unbroken and broken regimes \cite{SupplementalMaterial}. In other words, the difference between incoming and outgoing pseudophoton flux must cancel out, such that $\mathcal{Q}=\sum_n \mathcal{Q}_n$ vanishes, with
    \begin{equation} \label{eq:difference}
       \mathcal{Q}_n=\operatorname{sgn}(\omega_n)\left[ |\Psi_n^\mathrm{out}|^2 - |\Psi_n^\mathrm{in}|^2 \right].
    \end{equation}
    Thus, the pseudophoton flux difference can cancel out either because $|\Psi_n^\mathrm{out}| = |\Psi_n^\mathrm{in}|$, or because the signature term carefully balances the contribution of each channel. Panel (d) shows the pseudophoton flux difference \eqref{eq:difference} for each eigenvector component (blue dots, connected by a guide-to-the-eye line). As can be seen from our results, while the conservation of pseudophoton flux is satisfied in both regimes, with $\mathcal{Q}=\sum_n \mathcal{Q}_n=0$, the components $\mathcal{Q}_n$ exhibit completely different behavior in the unbroken and broken regimes. Specifically, while all the components $\mathcal{Q}_n$ vanish in the unbroken regime, such that $Q$ trivially vanishes, some become non-zero in the broken regime ($n=0,-1$) as a consequence of spontaneous symmetry breaking. 
    Importantly, however, the total flux difference always satisfies $\sum_n Q_n=0$.
    This is a result of the careful balance between positive and negative frequency components [see Eq.~\eqref{eq:difference}] leading to the conservation of pseudophoton flux for this dispersive and lossless time-modulated scatterer, even in the symmetry broken regime.

    \subsection{Experimental observation of spontaneous symmetry breaking}

    We realize this parametric resonator with a microstrip transmission line side-coupled to a ring resonator loaded with two varactor diodes, [see Fig.~\ref{fig:1}(c)], whose capacitance (and hence the resonance frequency of the ring) is modulated by an applied time-periodic voltage. As shown in the SM \cite{SupplementalMaterial}, this network is the one-dimensional analogue of the time-modulated point dipole of Eq.~\eqref{eq:parametric}; see Methods for details on the sample and measurements. 

    The pseudounitarity of the Floquet scattering matrix discussed above relies on the absence of material damping. The losses inevitably present in any experimental realization therefore mean that Eq.~\eqref{eq:pseudounitarity} no longer holds exactly, just as they render the scattering matrix of a static system sub-unitary. Signatures of pseudounitarity and of the symmetry-breaking transition nonetheless remain clearly observable.
    Moreover, the balance between radiative and material damping allows us to distinguish two regimes: the overcoupled regime, where radiative loss dominates, and the undercoupled regime, where material loss prevails.
    In the remainder of this work, we focus on the overcoupled scenario, as it is closer to the lossless limit, in which the ratio of ohmic to radiative losses tends to zero.
    The results for the undercoupled regime are provided in the SM \cite{SupplementalMaterial}.

    Figure \ref{fig:experiment} illustrates the behavior of the overcoupled microwave resonator through the eigenvalues and normalized right eigenvectors of its Floquet scattering matrix, $\Psi_r^{\mathrm{in}}$.
    In Fig.~\ref{fig:experiment}(a)-(b), we show the experimentally obtained eigenvalues of $\mathrm{S}_\mathrm{F}$ in the complex plane as a function of the amplitude of an external time-periodic voltage applied to the ring resonator, which here acts as the modulation strength, for $\omega=0.5000\,\Omega$ (a) and $0.5003\,\Omega$ (b).
    The modulation frequency $\Omega=2\,\omega_0$ is fixed at twice the resonance frequency, with $\omega_0=300\,$MHz.  

    In both cases, the eigenvalues associated to the symmetry breaking transition are inside the unit circle at small modulation voltages (blue/beige dots), signaling sub-unitary scattering induced by material damping.
    At the parametric resonance condition, see panel (a), as the modulation strength increases, two eigenvalues approach one another until they coalesce at an EP, marking the onset of spontaneous symmetry breaking. Upon further increasing the modulation strength, the eigenvalues drift apart along the real axis. 
    This shows that the system undergoes spontaneous symmetry breaking at the parametric resonance condition, and that the system exhibits CPA or lasing. 
    Importantly, however, material damping removes the degeneracy between these two phenomena, which arise here at two different modulation amplitude thresholds. 
    As shown later on, this allows us to experimentally observe the CPA condition, while avoiding the exponential amplification of noise induced by lasing. 

    The behavior away from the parametric resonance condition is, however, qualitatively different. 
    In this case, shown in panel (b), the eigenvalues of the Floquet scattering matrix show an avoided crossing upon increasing the modulation amplitude.
    This is a consequence of the losses present in the experiment, and the same features are also reproduced by the theoretical model when material damping is included (see SM \cite{SupplementalMaterial}). 
    Note that the symmetry-breaking transition itself is not confined to the parametric resonance condition: in the lossless model, EPs also occur for $\omega \neq \Omega/2$ [see Figs.~\ref{fig:2}(b),(c)]. It is the material damping that lifts them into avoided crossings.

    To confirm the existence of a true EP at the parametric resonance condition, we evaluate the Petermann factor $K$ of the complete set of eigenvectors of the scattering matrix \cite{Wiersig2023}.
    This quantity measures the sensitivity of an eigenstate to perturbations arising from the non-orthogonality of the eigenvectors. 
    It is equal to unity for orthogonal eigenvectors and diverges as two eigenvectors coalesce, thereby providing a clear signature of an EP.
    In Fig.~\ref{fig:experiment}(c), we plot the maximum Petermann factor of the scattering matrix as a function of the modulation amplitude for input frequencies $\omega=0.5000\,\Omega$, $0.5003\,\Omega$, and $0.5017\,\Omega$. 
    The Petermann factor exhibits a pronounced peak at the parametric resonance condition (blue line), confirming the coalescence of eigenvectors. By contrast, as the input frequency is detuned away from the parametric resonance condition, the peak rapidly disappears, indicating the emergence of an avoided crossing.
    

    Interestingly, the fact that the EP survives the presence of loss in our experiment is a consequence of an additional symmetry of the system: reflection–conjugation (RC) symmetry.
    Under this symmetry, the Floquet modes are invariant under the combined operation of frequency reflection, $\omega \rightarrow -\omega$, and complex conjugation \cite{Calvin2025}.
    This symmetry holds at $\omega=\Omega/2$, even in the presence of dispersion and loss, and is reflected in the eigenvectors of the system, as shown in Fig.~\ref{fig:experiment}(d).
    In particular, we observe that in the broken regime (right panels), eigenvector 2 remains invariant under RC symmetry, whereas eigenvector 4, taken away from the parametric resonance condition, does not exhibit this symmetry. 
    RC symmetry moreover ensures the presence of multiple EPs, one for each parametric resonance $\Omega=2 \,\omega_0/n$, with $n$ an integer (see SM \cite{SupplementalMaterial}).
    On the other hand, while time-reversal symmetry is formally broken by the material loss of the resonator, we observe that the phases of the eigenstates in the unbroken regime (left panels) are almost time-reversal invariant.
    This is not the case in the broken regime, just as it happened in the lossless scenario.
    
    Figure \ref{fig:experiment}(d) also shows the measured values of the Floquet components of the pseudophoton flux difference $\mathcal{Q}_n$ [see Eq.~\eqref{eq:difference}].
    Importantly, the breaking of pseudounitarity by material damping leads to violation of pseudophoton flux conservation, so that $\mathcal{Q} \neq 0$.
    Interestingly, however, eigenstates in the broken regime (see starred eigenvalues 2 and 4) nevertheless maintain a balanced distribution of pseudophoton flux difference, so that $\mathcal{Q} \simeq 0$. 
    This robustness is a signature of the underlying pseudounitary scattering, which, although it does not formally hold in the presence of loss, still manifests itself in the eigenstate structure.
    As shown by Eq.~\eqref{eq:difference}, the pseudophoton flux can cancel out either because the incoming and outgoing fields are equal, or thanks to the signature providing a balanced total flux. 
    As discussed above, in the lossless case, it is the first mechanism that is responsible for $\mathcal{Q}=0$ in the unbroken regime, while it is the second one that plays a role in the broken regime, by precisely balancing the contribution of positive and negative frequency channels. This balanced contribution of positive and negative frequency channels in the broken regime is maintained in the presence of loss (see Fig.~\ref{fig:experiment}).  
    Importantly, this shows that even in a realistic experiment, pseudounitarity is revealed by inducing a vanishing pseudophoton flux difference $\mathcal{Q}$, in the spontaneously broken regime. At parametric resonance (eigenvector 2) $\mathcal{Q}$ vanishes identically, protected by RC symmetry. Away from parametric resonance (eigenvector 4), where this protection is absent and the exceptional point is replaced by an avoided crossing, the flux is nonetheless still almost perfectly balanced ($|\mathcal{Q}|=0.01$, much smaller than its values in the unbroken regime). This reveals pseudounitarity as a general fundamental property of realistic time-varying systems.

    \begin{figure}[t!]
        \includegraphics[width=1.00\linewidth]{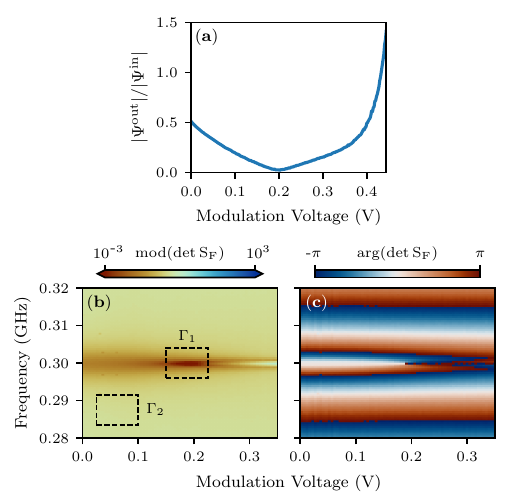}
        \caption{\textbf{Experimental observation of coherent perfect absorption.} Norm of the field scattered by the resonator (a) when excited by the CPA eigenvector as a function of the  amplitude of the applied time-modulated voltage, i.e., the modulation strength, and measurements of both the modulus (b) and argument (c) of the determinant of the Floquet scattering matrix, as a function of the modulation strength and the input frequency. 
        }
        \label{fig:4}
    \end{figure}
        
    \subsection{Coherent Perfect Absorption}
    
    Finally, we demonstrate that the system exhibits CPA in the vicinity of the small but finite eigenvalue depicted in Fig.~\ref{fig:experiment}(a) (see the starred eigenvalue 2). 
    We show in Fig.~\ref{fig:4}(a) the measured norm of the field scattered by the system when excited by the CPA eigenvector as a function of the modulation voltage, normalised to the input field.
    At the modulation voltage corresponding to the CPA condition the scattered field is reduced by a factor of 45.5 compared to the input field, confirming near perfect absorption.
    
    To prove unequivocally that such vanishing absorption corresponds to CPA, we rely on the winding number of the Floquet scattering matrix \cite{Guo2023,Globosits2026} 
    \begin{equation}\label{eq:winding}
        \operatorname{wind}(\mathrm{S}_\mathrm{F},\Gamma)=\frac{1}{2\pi i}\int_\Gamma \dd{\phi} \partial_\phi \log[\det(\mathrm{S}_\mathrm{F})].
    \end{equation}
    Here, $\phi$ is any given tunable parameter, and the winding number equals $\pm\;1$ when the loop $\Gamma$ encloses a zero or a pole of $\det(\mathrm{S}_\mathrm{F})$.
    We present in Fig.~\ref{fig:4}(b)-(c) the experimentally measured modulus and argument of the determinant of the scattering matrix as a function of modulation voltage and input frequency.
    Using these measurements, we numerically evaluate the winding number along the loop $\Gamma_{1}$ which encircles a zero of the scattering matrix visible at the parametric resonance condition $\omega=\Omega/2$, and find $\operatorname{wind}(\mathrm{S}_\mathrm{F},\Gamma_1)=1.0167-0.0376i$.
    This value close to $1$ verifies the existence of a CPA point in the system.
    For comparison, we also evaluate the winding number along the loop $\Gamma_2$ that does not encircle a zero, and obtain a very small value for $\operatorname{wind}(\mathrm{S}_\mathrm{F},\Gamma_2)=0.0001-0.0036i$, as expected.

\section{Conclusions}
We have observed a spontaneous symmetry-breaking transition in the Floquet scattering matrix $\mathrm{S}_\mathrm{F}$ of a periodically driven open resonator, tracking its eigenvalues from an unbroken regime through an exceptional point into a
symmetry-broken regime that culminates, at parametric resonance, in coherent perfect absorption. Because this transition follows from the algebraic structure of $\mathrm{S}_\mathrm{F}$ alone, the same physics should arise in any spatially finite, periodically driven scattering system. Crucially, our measurements show that it survives in a realistic experiment: material damping
formally invalidates pseudounitarity, yet the symmetry-broken eigenstates retain an almost perfectly balanced pseudophoton flux, so that the underlying conservation law remains visible in the eigenvector structure. Pseudounitarity is thus a robust organizing principle for realistic time-varying systems, not an idealization confined to the lossless limit.

Since the relevant scale is the ratio of the modulation to the resonance frequency rather than any absolute one, we expect these results to carry over to modulated dipoles, metatoms, and plasmonic nanoparticles \cite{Ptitcyn2019,Mirmoosa2022,BlancoDePaz2025,Verde2026}, to plasmonic platforms in the terahertz domain \cite{guo2025plasmonic}, and, beyond electromagnetism, to driven phononic \cite{Basini2026} and acoustic \cite{tong2025observation,liu2026temporal} systems -- offering a common
framework for the coherent wave control, perfect absorption, and amplification already reported at optical frequencies
\cite{galiffi2023broadband,galiffi_optical_2026}. More broadly, the exceptional points and the coherent perfect absorption reached here require no balance of gain and loss, but only the temporal driving of a passive structure. Combined with the ability to shape the multispectral input state, this suggests routes towards dynamically reconfigurable absorbers, parametric amplifiers operated at
a symmetry-breaking threshold, and exceptional-point sensing in which the modulation itself serves as the tuning parameter.




\section{Methods}

\subsection{Theory}
\subsubsection{Floquet Scattering Matrix of the Parametric Dipole}

    In the following, we provide details on the derivation of the Floquet scattering matrix $\mathrm{S}_\mathrm{F}$ of the three-dimensional time-varying dipole introduced in Eq.~\eqref{eq:parametric}. We employ the electromagnetic Green-tensor method together with Floquet theory. Specifically, the Floquet scattering matrix is given by
    \begin{equation*}
        \mathrm{S}_\mathrm{F}(\omega)=\mathbb{I}+2\mathrm{G}_\mathrm{\Im}(\omega)\mathrm{A}(\omega).
    \end{equation*}
    Here, we introduced the Floquet dressed dipole polarizability, that incorporates retardation effects, as
    \begin{equation*}
    \mathrm{A}(\omega)=\left[\mathrm{A}_\mathrm{B}^{-1}(\omega)-\mathrm{G}_\mathrm{\Im}(\omega)\right]^{-1},
    \end{equation*}
    and the imaginary part of the free-space Green tensor in Floquet space given by
    \begin{equation*}
            \mathrm{G}_\mathrm{\Im}(\omega)|_{nm}=i\delta_{nm}\frac{\omega_n^3}{6\pi\varepsilon_0c^3},
    \end{equation*}
    with $\delta_{nm}$ the Kronecker delta. Furthermore, the Floquet bare dipole polarizability reads
    \begin{equation*}
        \mathrm{A}_\mathrm{B}(\omega) = \kappa \left[ \mathrm{A}_\mathrm{0}^{-1}(\omega) + \mathrm{M} \right]^{-1},
    \end{equation*}
    where we made us of the matrices
    \begin{equation*}
            \mathrm{M}_{nm} = \mathrm{W}_{n-m} [1-\delta_{nm}],
    \end{equation*}
    and
    \begin{equation*}
            \mathrm{A}^{-1}_{0}(\omega)|_{nm} = \delta_{nm} \left[ \mathrm{W}_{0} - \omega_n^2 - i\gamma\omega_n \right].
    \end{equation*}
    In our simulations, we compute the Floquet scattering matrix using $N_F=10$ Floquet modes, which makes it a $10 \times 10$ matrix. More details can be found in the SM \cite{SupplementalMaterial}. 

\subsubsection{One-dimensional Scattering Problem}

    A capacitively coupled resonator attached to a transmission line provides a one-dimensional realization of the periodically driven dipole scattering problem. The telegrapher's equations are formally equivalent to the one-dimensional Maxwell equations, with the resonator acting as a shunt admittance that introduces a discontinuity in the current at the coupling point, analogous to the discontinuity in the magnetic field produced by the electric dipole. Consequently, the circuit scattering matrix obeys the same pseudounitary symmetry as the dipole, with the conserved weighted flux
    \begin{equation*}
        \tilde{\mathrm{S}}(\omega) = \sum_n \omega_n^{-1} \left[ V_n(\omega)I^*_n(\omega) + V^*_n(\omega)I_n(\omega) \right].
    \end{equation*}
    where $V_n$ and $I_n$ denote the Floquet components of voltage and current amplitudes, respectively.
    Note that in the three dimensional scattering dipole problem there is a single spatial port for each Floquet mode, while the one dimensional scattering setup has two ports per Floquet mode: one for left-incoming and one for right-incoming waves. Due to the mirror symmetry of the setup, however, the components of the eigenvectors corresponding to the waves incoming from the left and right are identical. Thus, in Fig.~\ref{fig:experiment}(d) we only show half of the components of the respective Floquet eigenvectors corresponding to the incoming waves from the right port.

\subsection{Experiments}

\subsubsection{The sample}

The sample consists of a ring resonator incorporating two varactor diodes (Infineon BBY55-03W) coupled to a microstrip transmission line, see Fig.~\ref{fig:1}(c). The sample was fabricated using a LPKF ProtoLaser H4 from a double-sided 1.5 mm thick Astra MT-77 PCB substrate with 18 \unit{\micro\meter} thick copper layers. All track widths are 3.78 mm, resulting in a 50 $\Omega$ impedance for the transmission line. The transmission line is 150 mm long, and the ring resonator has a length of 120 mm and a width of 10 mm. 1 mm gaps in the ring resonator are placed equidistant from the ends of the resonator, across which the diodes are soldered. For the experiment described here the gap between the transmission line and the resonator is 100 \unit{\micro\meter}, ensuring that the resonator is in the overcoupled regime. Bias lines attached across the diodes allow their capacitance to be modified, changing the resonance frequency of the resonator -- whether statically with a DC bias voltage, or dynamically with an additional temporal modulation. For characterisation of the sample, showing the change in resonance frequency as a function of applied DC bias voltage, see the SM~\cite{SupplementalMaterial}. 

\subsubsection{Measurements of the Floquet scattering matrices}

An offset bias voltage of 5.66 V was chosen to place the (static) resonance frequency, $\omega_0$, at 0.3 GHz. On top of this a sinusoidal modulating voltage was applied of frequency $\Omega=2\omega_0$ and amplitude $\Delta$. We construct our Floquet scattering matrices in the Fourier basis by injecting successive sinusoidal and co-sinusoidal waveforms of the requisite frequencies from either side of our transmission line using a Keysight M8195A arbitrary waveform generator and measuring the requisite reflected and transmitted spectral components using a Keysight DSO-Z-254A  oscilloscope. After undertaking a post-processing calibration step these measurements are used to populate our scattering matrices, and we undertake this for varying modulation amplitudes. In our experiments we use $N_F=4$, resulting in $2N_F\times2N_F$ Floquet scattering matrices. For more details of the experimental set-up, measurement and calibration processes see the SM~\cite{SupplementalMaterial}.

\section{Data Availability}

All data and codes are available from the University of Exeter open research repository, DOI:(to be added at proof stage after deposit).

\section{Acknowledgements}

I.R.H.~and S.A.R.H.~acknowledge support from the EPSRC via the META4D Programme Grant (EP/Y015673/1). M.V., T.F.A.~and P.A.H.~acknowledge funding from the European Union through the ERC grant TIMELIGHT under GA101115792 and from the Spanish Ministry for Science, Innovation, and Universities – Agencia Estatal de Investagación (AEI) through Grants No.~PID2021-125894NB-I00, No.~CEX2018-000805-M (through the María de Maeztu program for Units of Excellence in Research and Developments) and Grant No.~RYC2021-031568-I (Ramón y Cajal program).

\section{Author Contributions.}

IRH, DG, PAH, and SR initiated the project. MV, TFA, DG, SR, SARH, and  PAH developed the theory. IRH designed and performed the experiments. IRH, SARH, and MV analysed the data. CH and SARH contributed towards the understanding of RC symmetry breaking. MV, TFA, DG, PAH, SR, and IRH wrote the paper, with input from all the authors. All authors interpreted the results. PAH, SR and SARH supervised the project.

\bibliographystyle{apsrev4-2}
\bibliography{bib}

@article{El-Ganainy2018,
author={El-Ganainy, Ramy
and Makris, Konstantinos G.
and Khajavikhan, Mercedeh
and Musslimani, Ziad H.
and Rotter, Stefan
and Christodoulides, Demetrios N.},
title={Non-Hermitian physics and PT symmetry},
journal={Nat. Phys.},
year={2018},
month={Jan},
day={01},
volume={14},
number={1},
pages={11-19},
issn={1745-2481},
doi={10.1038/nphys4323},
url={https://doi.org/10.1038/nphys4323}
}

@article{Ozdemir2019,
author={{\"O}zdemir, {\c{S}} K.
and Rotter, S.
and Nori, F.
and Yang, L.},
title={Parity--time symmetry and exceptional points in photonics},
journal={Nat. Mater.},
year={2019},
month={Aug},
day={01},
volume={18},
number={8},
pages={783-798},
issn={1476-4660},
doi={10.1038/s41563-019-0304-9},
url={https://doi.org/10.1038/s41563-019-0304-9}
}

@article{Chong2010,
  title = {Coherent Perfect Absorbers: Time-Reversed Lasers},
  author = {Chong, Y. D. and Ge, Li and Cao, Hui and Stone, A. D.},
  journal = {Phys. Rev. Lett.},
  volume = {105},
  issue = {5},
  pages = {053901},
  numpages = {4},
  year = {2010},
  month = {Jul},
  publisher = {American Physical Society},
  doi = {10.1103/PhysRevLett.105.053901},
}

@article{Wan2011,
author = {Wenjie Wan  and Yidong Chong  and Li Ge  and Heeso Noh  and A. Douglas Stone  and Hui Cao },
title = {Time-Reversed Lasing and Interferometric Control of Absorption},
journal = {Science},
volume = {331},
number = {6019},
pages = {889-892},
year = {2011},
doi = {10.1126/science.1200735},
URL = {https://www.science.org/doi/abs/10.1126/science.1200735},
}

@article{Galiffi2022,
    author = {Emanuele Galiffi and Romain Tirole and Shixiong Yin and Huanan Li and Stefano Vezzoli and Paloma A. Huidobro and M{\'a}rio G. Silveirinha and Riccardo Sapienza and Andrea Al{\`u} and J. B. Pendry},
    title = {{Photonics of time-varying media}},
    volume = {4},
    journal = {Adv. Photonics},
    number = {1},
    publisher = {SPIE},
    pages = {014002},
    year = {2022},
    doi = {10.1117/1.AP.4.1.014002},
    URL = {https://doi.org/10.1117/1.AP.4.1.014002}
}

@article{Engheta2023,
    author = {Nader Engheta },
    title = {Four-dimensional optics using time-varying metamaterials},
    journal = {Science},
    volume = {379},
    number = {6638},
    pages = {1190-1191},
    year = {2023},
    doi = {10.1126/science.adf1094},
    URL = {https://www.science.org/doi/abs/10.1126/science.adf1094},
}

@article{Lyubarov2022,
    author = {Mark Lyubarov  and Yaakov Lumer  and Alex Dikopoltsev  and Eran Lustig  and Yonatan Sharabi  and Mordechai Segev },
    title = {Amplified emission and lasing in photonic time crystals},
    journal = {Science},
    volume = {377},
    number = {6604},
    pages = {425-428},
    year = {2022},
    doi = {10.1126/science.abo3324},
    URL = {https://www.science.org/doi/abs/10.1126/science.abo3324},
}

@article{Globosits2024,
  title = {Pseudounitary Floquet scattering matrix for wave-front shaping in time-periodic photonic media},
  author = {Globosits, David and H\"upfl, Jakob and Rotter, Stefan},
  journal = {Phys. Rev. A},
  volume = {110},
  issue = {5},
  pages = {053515},
  numpages = {17},
  year = {2024},
  month = {Nov},
  publisher = {American Physical Society},
  doi = {10.1103/PhysRevA.110.053515},
  url = {https://link.aps.org/doi/10.1103/PhysRevA.110.053515}
}

@Article{Globosits2026,
author={Globosits, David
and Garg, Puneet
and H{\"u}pfl, Jakob
and Can{\'o}s Valero, Adri{\`a}
and Weiss, Thomas
and Rockstuhl, Carsten
and Rotter, Stefan},
title={Exceptional points, lasing, and coherent perfect absorption in Floquet scattering systems},
journal={Light Sci. Appl.},
year={2026},
month={Sep},
day={09},
volume={15},
number={1},
pages={375},
issn={2047-7538},
doi={10.1038/s41377-026-02380-9},
url={https://doi.org/10.1038/s41377-026-02380-9}
}

@article{Pendry2023,
author = {J. B. Pendry},
journal = {Opt. Express},
number = {1},
pages = {452--458},
publisher = {Optica Publishing Group},
title = {Photon number conservation in time dependent systems \[Invited\]},
volume = {31},
month = {Jan},
year = {2023},
url = {https://opg.optica.org/oe/abstract.cfm?URI=oe-31-1-452},
doi = {10.1364/OE.476961},
}

@article{Floquet1888,
  author = {Floquet, Gaston},
  title = {Sur les équations différentielles linéaires à coefficients périodiques},
  journal = {Ann. Sci. Éc. Norm. Supér.},
  publisher = {Elsevier},
  volume = {2e série, 12},
  pages = {47--88},
  year = {1883},
  doi = {10.24033/asens.220},
  url = {https://numdam.org}
}

@misc{SupplementalMaterial,
    title={{See Supplemental Material at [URL] for details on our formalism and complementary results.}},
}

@book{Newton2013,
  title={Scattering theory of waves and particles},
  author={Newton, Roger G},
  year={2013},
  publisher={Springer Science \& Business Media}
}

@book{christodoulides2018parity,
  title={Parity-time symmetry and its applications},
  author={Christodoulides, Demetrios and Yang, Jianke and others},
  volume={280},
  year={2018},
  publisher={Springer}
}

@article{Mostafazadeh2001,
    author = {Mostafazadeh, Ali},
    title = {Pseudounitary operators and pseudounitary quantum dynamics},
    journal = {J. Math. Phys.},
    volume = {45},
    number = {3},
    pages = {932-946},
    year = {2004},
    month = {03},
    issn = {0022-2488},
    doi = {10.1063/1.1646448},
    url = {https://doi.org/10.1063/1.1646448},
}

@article{Ruter2010,
author={R{\"u}ter, Christian E.
and Makris, Konstantinos G.
and El-Ganainy, Ramy
and Christodoulides, Demetrios N.
and Segev, Mordechai
and Kip, Detlef},
title={Observation of parity--time symmetry in optics},
journal={Nat. Phys.},
year={2010},
month={Mar},
day={01},
volume={6},
number={3},
pages={192-195},
issn={1745-2481},
doi={10.1038/nphys1515},
url={https://doi.org/10.1038/nphys1515}
}

@article{Calvin2025,
  title = {Symmetry-protected lossless modes in dispersive time-varying media},
  author = {Hooper, Calvin M. and Capers, James R. and Hooper, Ian R. and Horsley, Simon A. R.},
  journal = {Phys. Rev. A},
  volume = {111},
  issue = {3},
  pages = {033507},
  numpages = {10},
  year = {2025},
  month = {Mar},
  publisher = {American Physical Society},
  doi = {10.1103/PhysRevA.111.033507},
  url = {https://link.aps.org/doi/10.1103/PhysRevA.111.033507}
}

@article{Guo2023,
  title = {Singular topology of scattering matrices},
  author = {Guo, Cheng and Li, Jiazheng and Xiao, Meng and Fan, Shanhui},
  journal = {Phys. Rev. B},
  volume = {108},
  issue = {15},
  pages = {155418},
  numpages = {21},
  year = {2023},
  month = {Oct},
  publisher = {American Physical Society},
  doi = {10.1103/PhysRevB.108.155418},
  url = {https://link.aps.org/doi/10.1103/PhysRevB.108.155418}
}

@article{Wiersig2023,
  title = {Petermann factors and phase rigidities near exceptional points},
  author = {Wiersig, Jan},
  journal = {Phys. Rev. Res.},
  volume = {5},
  issue = {3},
  pages = {033042},
  numpages = {9},
  year = {2023},
  month = {Jul},
  publisher = {American Physical Society},
  doi = {10.1103/PhysRevResearch.5.033042},
  url = {https://link.aps.org/doi/10.1103/PhysRevResearch.5.033042}
}

@article{nasari_observation_2026,
	title = {Observation of {Floquet} rotational super-radiance},
	volume = {655},
	issn = {1476-4687},
	url = {https://doi.org/10.1038/s41586-026-10725-y},
	doi = {10.1038/s41586-026-10725-y},
	number = {8123},
	journal = {Nature},
	author = {Nasari, Hadiseh and Moussa, Hady and Kasahara, Yoshiaki and Thielens, Arno and Alù, Andrea},
	month = jul,
	year = {2026},
	pages = {608--616},
}

@article{bacot_time_2016,
	title = {Time reversal and holography with spacetime transformations},
	volume = {12},
	issn = {1745-2481},
	url = {https://doi.org/10.1038/nphys3810},
	doi = {10.1038/nphys3810},
	number = {10},
	journal = {Nat. Phys.},
	author = {Bacot, Vincent and Labousse, Matthieu and Eddi, Antonin and Fink, Mathias and Fort, Emmanuel},
	month = oct,
	year = {2016},
	pages = {972--977},
}

@article{galiffi2023broadband,
  title={Broadband coherent wave control through photonic collisions at time interfaces},
  author={Galiffi, Emanuele and Xu, Gengyu and Yin, Shixiong and Moussa, Hady and Ra’di, Younes and Al{\`u}, Andrea},
  journal={Nat. Phys.},
  volume={19},
  number={11},
  pages={1703--1708},
  year={2023},
  publisher={Nature Publishing Group UK London},
  doi = {https://doi.org/10.1038/s41567-023-02165-6}
}

@article{moussa2023observation,
  title={Observation of temporal reflection and broadband frequency translation at photonic time interfaces},
  author={Moussa, Hady and Xu, Gengyu and Yin, Shixiong and Galiffi, Emanuele and Ra’di, Younes and Al{\`u}, Andrea},
  journal={Nat. Phys.},
  volume={19},
  number={6},
  pages={863--868},
  year={2023},
  publisher={Nature Publishing Group UK London},
  doi = {https://doi.org/10.1038/s41567-023-01975-y}
}

@article{wang2025expanding,
  title={Expanding momentum bandgaps in photonic time crystals through resonances},
  author={Wang, X and Garg, P and Mirmoosa, MS and Lamprianidis, AG and Rockstuhl, C and Asadchy, VS},
  journal={Nat. Photonics},
  volume={19},
  number={2},
  pages={149--155},
  year={2025},
  publisher={Nature Publishing Group UK London},
  doi = {https://doi.org/10.1038/s41566-024-01563-3}
}

@article{buddhiraju2020photonic,
  title={Photonic refrigeration from time-modulated thermal emission},
  author={Buddhiraju, Siddharth and Li, Wei and Fan, Shanhui},
  journal={Phys. Rev. Lett.},
  volume={124},
  number={7},
  pages={077402},
  year={2020},
  publisher={APS},
  doi = {https://doi.org/10.1103/PhysRevLett.124.077402}
}

@article{vezzoli2018optical,
  title={Optical time reversal from time-dependent epsilon-near-zero media},
  author={Vezzoli, Stefano and Bruno, Vincenzo and DeVault, Clayton and Roger, Thomas and Shalaev, Vladimir M and Boltasseva, Alexandra and Ferrera, Marcello and Clerici, Matteo and Dubietis, Audrius and Faccio, Daniele},
  journal={Phys. Rev. Lett.},
  volume={120},
  number={4},
  pages={043902},
  year={2018},
  publisher={APS},
  doi = {https://doi.org/10.1103/PhysRevLett.120.043902}
}

@article{lee2026analogs,
  title={Analogs of spontaneous emission and lasing in photonic time crystals},
  author={Lee, Kyungmin and Kyung, Minwook and Kim, Yung and Park, Jagang and Lee, Hansuek and Choi, Joonhee and Chan, CT and Shin, Jonghwa and Kim, Kun Woo and Min, Bumki},
  journal={Phys. Rev. Lett.},
  volume={136},
  number={9},
  pages={093802},
  year={2026},
  publisher={APS},
  doi = {https://doi.org/10.1103/hh9h-qzpk}
}

@article{boltasseva2024photonic,
  title={Photonic time crystals: from fundamental insights to novel applications: opinion},
  author={Boltasseva, A and Shalaev, VM and Segev, M},
  journal={Opt. Mater. Express},
  volume={14},
  number={3},
  pages={592--597},
  year={2024},
  publisher={Optica Publishing Group},
  doi = {https://doi.org/10.1364/OME.511801}
}

@article{xiong2025observation,
  title={Observation of wave amplification and temporal topological state in a non-synthetic photonic time crystal},
  author={Xiong, Jiang and Zhang, Xudong and Duan, Longji and Wang, Jiarui and Long, Yang and Hou, Haonan and Yu, Letian and Zou, Linyang and Zhang, Baile},
  journal={Nat. Comm.},
  volume={16},
  number={1},
  pages={11182},
  year={2025},
  publisher={Nature Publishing Group UK London},
  doi = {https://doi.org/10.1038/s41467-025-66154-4}
}

@article{reyes2015observation,
  title={Observation of genuine wave vector (k or $\beta$) gap in a dynamic transmission line and temporal photonic crystals},
  author={Reyes-Ayona, JR and Halevi, P},
  journal={Appl. Phys. Lett.},
  volume={107},
  number={7},
  year={2015},
  publisher={AIP Publishing},
  url = {https://doi.org/10.1063/1.4928659}
}

@article{yu-2009,
	author = {Yu, Zongfu and Fan, Shanhui},
	journal = {Nat. Photonics},
	month = {1},
	number = {2},
	pages = {91--94},
	title = {{Complete optical isolation created by indirect interband photonic transitions}},
	volume = {3},
	year = {2009},
	url = {https://doi.org/10.1038/nphoton.2008.273},
}

@article{tirole,
	author = {Tirole, Romain and Vezzoli, Stefano and Galiffi, Emanuele and Robertson, Iain and Maurice, Dries and Tilmann, Benjamin and Maier, Stefan A. and Pendry, John B. and Sapienza, Riccardo},
	journal = {Nat. Phys.},
	month = {4},
	number = {7},
	pages = {999--1002},
	title = {{Double-slit time diffraction at optical frequencies}},
	volume = {19},
	year = {2023},
	url = {https://doi.org/10.1038/s41567-023-01993-w},
}

@article{guo2025plasmonic,
  author = {Guo, Tingwen and Sueiro, Jules and Andolina, Gian Marcello and Levchuk, Artem and Ponzoni, Stefano and Grasset, Romain and Monthe, Donald and Aupiais, Ian and Daineka, Dmitri and Briatico, Javier and de Oliveira, Thales VAG and Ponomaryov, Alexey and Arshad, Atiqa and Karimbana-Kandy, Arjun and Prajapati, Gulloo Lal and Ilyakov, Igor and Deinert, Jan-Christoph and Maehrlein, Sebastian F. and Perfetti, Luca and Schirò, Marco and Laplace, Yannis},
  title = {Plasmonic metamaterial time crystal},
  journal = {Nature},
  year = {2026},
  volume = {656},
  number = {8127},
  pages = {343--348},
  doi = {10.1038/s41586-026-10825-9},
  url = {https://doi.org/10.1038/s41586-026-10825-9},
  isbn = {1476-4687}
}

@article{CanosValero2026,
author = {Canós Valero, Adrià and Gladyshev, Sergei and Globosits, David and Rotter, Stefan and Muljarov, Egor A. and Weiss, Thomas},
title = {Revealing the Resonant Physics of Open Photonic Time Crystals},
journal = {Laser Photonics Rev.},
volume = {e71318},
url = {https://doi.org/10.1002/lpor.71318},
year = {2026}
}

@article{Ptitcyn2022,
	author = {Ptitcyn, Grigorii and Lamprianidis, Aristeidis and Karamanos, Theodosios and Asadchy, Viktar and Alaee, Rasoul and Müller, Marvin and Albooyeh, Mohammad and Mirmoosa, Mohammad Sajjad and Fan, Shanhui and Tretyakov, Sergei and Rockstuhl, Carsten},
	journal = {Laser Photonics Rev.},
	month = {12},
	number = {3},
	title = {{Floquet–Mie Theory for Time‐Varying Dispersive Spheres}},
	volume = {17},
	year = {2022},
	url = {https://doi.org/10.1002/lpor.202100683},
}

@article{wang2023metasurface,
  title={Metasurface-based realization of photonic time crystals},
  author={Wang, Xuchen and Mirmoosa, Mohammad Sajjad and Asadchy, Viktar S and Rockstuhl, Carsten and Fan, Shanhui and Tretyakov, Sergei A},
  journal={Sci. Adv.},
  volume={9},
  number={14},
  pages={eadg7541},
  year={2023},
  publisher={American Association for the Advancement of Science},
  doi = {10.1126/sciadv.adg7541}
}

@article{garg2025photonic,
author = {Puneet Garg  and Evangelos Almpanis  and Leander Zimmer  and Jan David Fischbach  and Xuchen Wang  and Mohammad S. Mirmoosa  and Markus Nyman  and Nikolaos Stefanou  and Nikolaos Papanikolaou  and Viktar Asadchy  and Carsten Rockstuhl },
title = {Photonic time crystals assisted by quasi-bound states in the continuum},
journal = {Science Advances},
volume = {12},
number = {33},
pages = {eaed4055},
year = {2026},
doi = {10.1126/sciadv.aed4055},
URL = {https://www.science.org/doi/abs/10.1126/sciadv.aed4055}
}

@article{galiffi2020wood,
  title={Wood anomalies and surface-wave excitation with a time grating},
  author={Galiffi, Emanuele and Wang, Yao-Ting and Lim, Zhen and Pendry, John B and Al{\`u}, Andrea and Huidobro, Paloma A},
  journal={Phys. Rev. Lett.},
  volume={125},
  number={12},
  pages={127403},
  year={2020},
  doi = {10.1103/PhysRevLett.125.127403},
  publisher={APS}
}

@article{zurita2010resonances,
  title={Resonances in the optical response of a slab with time-periodic dielectric function $\varepsilon$ (t)},
  author={Zurita-Sánchez, Jorge R and Halevi, P},
  journal={Phys. Rev. A},
  volume={81},
  number={5},
  pages={053834},
  year={2010},
  publisher={APS},
  doi = {https://doi.org/10.1103/PhysRevA.81.053834}
}

@article{Mirmoosa2022,
	author = {Mirmoosa, M S and Koutserimpas, T T and Ptitcyn, G A and Tretyakov, S A and Fleury, R},
	journal = {New J. Phys.},
	month = {4},
	number = {6},
	pages = {063004},
	title = {{Dipole polarizability of time-varying particles}},
	volume = {24},
	year = {2022},
	url = {https://doi.org/10.1088/1367-2630/ac6b4c},
}

@article{Ptitcyn2019,
  title = {Time-modulated meta-atoms},
  author = {Ptitcyn, G. and Mirmoosa, M. S. and Tretyakov, S. A.},
  journal = {Phys. Rev. Res.},
  volume = {1},
  issue = {2},
  pages = {023014},
  numpages = {11},
  year = {2019},
  month = {Sep},
  publisher = {American Physical Society},
  doi = {10.1103/PhysRevResearch.1.023014},
  url = {https://link.aps.org/doi/10.1103/PhysRevResearch.1.023014}
}

@article{BlancoDePaz2025,
author = {Blanco de Paz, María and Deop-Ruano, Juan R. and Solís, Diego M. and Manjavacas, Alejandro},
title = {Lattice Resonances in Periodic Arrays of Time-Modulated Scatterers},
journal = {Laser Photonics Rev.},
volume = {20},
number = {17},
 year = {2026},
pages = {e03171},
doi = {https://doi.org/10.1002/lpor.202503171},
url = {https://onlinelibrary.wiley.com/doi/abs/10.1002/lpor.202503171}
}

@article{Verde2026,
  title = {Optical response by time-varying plasmonic nanoparticles},
  author = {Verde, Miguel and Huidobro, Paloma A.},
  journal = {Phys. Rev. Res.},
  volume = {8},
  issue = {2},
  pages = {023073},
  numpages = {12},
  year = {2026},
  month = {Apr},
  publisher = {American Physical Society},
  doi = {10.1103/l1xz-kz8w},
  url = {https://link.aps.org/doi/10.1103/l1xz-kz8w}
}

@article{estep_magnetic-free_2014,
	title = {Magnetic-free non-reciprocity and isolation based on parametrically modulated coupled-resonator loops},
	volume = {10},
	issn = {1745-2481},
	url = {https://doi.org/10.1038/nphys3134},
	doi = {10.1038/nphys3134},
	number = {12},
	journal = {Nat. Phys.},
	author = {Estep, Nicholas A. and Sounas, Dimitrios L. and Soric, Jason and Alù, Andrea},
	month = dec,
	year = {2014},
	pages = {923--927},
}

@article{
Pendry2008,
author = {J. B. Pendry },
title = {Time Reversal and Negative Refraction},
journal = {Science},
volume = {322},
number = {5898},
pages = {71-73},
year = {2008},
doi = {10.1126/science.1162087},
URL = {https://www.science.org/doi/abs/10.1126/science.1162087}}

@article{galiffi_optical_2026,
	title = {Optical coherent perfect absorption and amplification in a time-varying medium},
	volume = {20},
	issn = {1749-4893},
	url = {https://doi.org/10.1038/s41566-025-01833-8},
	doi = {10.1038/s41566-025-01833-8},
	number = {2},
	journal = {Nat. Photonics},
	author = {Galiffi, Emanuele and Harwood, Anthony C. and Vezzoli, Stefano and Tirole, Romain and Alù, Andrea and Sapienza, Riccardo},
	month = feb,
	year = {2026},
	pages = {163--169},
}

@article{Basini2026,
  title = {Terahertz-Driven Parametric Excitation of Raman-Active Phonons in ${\mathrm{LaAlO}}_{3}$},
  author = {Basini, M. and Unikandanunni, V. and Gabriele, F. and Cross, M. and Derrico, A. M. and Gray, A. X. and Hoffmann, M. C. and Forte, F. and Cuoco, M. and Bonetti, S.},
  journal = {Phys. Rev. Lett.},
  volume = {136},
  issue = {15},
  pages = {156902},
  numpages = {7},
  year = {2026},
  month = {Apr},
  publisher = {American Physical Society},
  doi = {10.1103/57p8-7mh9},
  url = {https://link.aps.org/doi/10.1103/57p8-7mh9}
}

@article{tong2025observation,
  title={Observation of momentum-band topology in PT-symmetric Floquet lattices},
  author={Tong, Shuaishuai and Zhang, Qicheng and Li, Gaohan and Zhang, Kun and Xie, Chun and Qiu, Chunyin},
  journal={Nat. Commun.},
  volume={16},
  number={1},
  pages={9975},
  year={2025},
  doi={10.1038/s41467-025-64915-9},
  publisher={Nature Publishing Group UK London}
}

@article{liu2026temporal,
  title={Temporal super-cell engineering and acoustic amplification in dispersive phononic time crystals},
  author={Liu, Ziling and Zhu, Xinghong and Zhang, Zhi-Guo and Zhang, Wei-Min and Chen, Xue and Yang, Yong-Qiang and Peng, Ruwen and Wang, Mu and Li, Jensen and Wu, Hong-Wei},
  journal={Nat. Commun.},
  year={2026},
  doi={10.1038/s41467-026-73459-5},
  publisher={Nature Publishing Group UK London},
  	volume = {17},
    number = {1},
    pages = {7079},
}

@article{Anderson1972, author={P. W. Anderson}, title={More Is Different},
  journal={Science}, volume={177}, pages={393}, year={1972},
  doi={10.1126/science.177.4047.393}}

@article{Beekman2019, author={A. J. Beekman and L. Rademaker and J. van Wezel},
  title={An Introduction to Spontaneous Symmetry Breaking},
  journal={SciPost Phys. Lect. Notes}, pages={11}, year={2019},
  doi={10.21468/SciPostPhysLectNotes.11}}

@book{Strocchi2008, author={F. Strocchi}, title={Symmetry Breaking},
  series={Lecture Notes in Physics}, volume={732}, edition={2},
  publisher={Springer}, address={Berlin}, year={2008}}

@article{Englert1964, author={F. Englert and R. Brout},
  title={Broken Symmetry and the Mass of Gauge Vector Mesons},
  journal={Phys. Rev. Lett.}, volume={13}, pages={321}, year={1964}, doi = {10.1103/PhysRevLett.13.321}}

@article{Higgs1964, author={P. W. Higgs},
  title={Broken Symmetries and the Masses of Gauge Bosons},
  journal={Phys. Rev. Lett.}, volume={13}, pages={508}, year={1964}, doi = {10.1103/PhysRevLett.13.508}}

@book{Chaikin1995, author={P. M. Chaikin and T. C. Lubensky},
  title={Principles of Condensed Matter Physics},
  publisher={Cambridge University Press}, year={1995}}

@article{CrossHohenberg1993, author={M. C. Cross and P. C. Hohenberg},
  title={Pattern formation outside of equilibrium},
  journal={Rev. Mod. Phys.}, volume={65}, pages={851}, year={1993}, doi = {10.1103/RevModPhys.65.851}}

@article{Buddhiraju2021, author={S. Buddhiraju and A. Dutt and M. Minkov
  and I. A. D. Williamson and S. Fan},
  title={Arbitrary linear transformations for photons in the frequency
  synthetic dimension},
  journal={Nat. Commun.}, volume={12}, pages={2401}, year={2021}, doi = {10.1038/s41467-021-22670-7}}

@article{Fan2022, author={L. Fan and Z. Zhao and K. Wang and A. Dutt and
  J. Wang and S. Buddhiraju and C. C. Wojcik and S. Fan},
  title={Multidimensional convolution operation with synthetic frequency
  dimensions in photonics},
  journal={Phys. Rev. Applied}, volume={18}, pages={034088}, year={2022}, doi = {10.1103/PhysRevApplied.18.034088}}

@article{Li1999,
  author  = {W. Li and L. E. Reichl},
  title   = {Floquet scattering through a time-periodic potential},
  journal = {Phys. Rev. B},
  volume  = {60},
  pages   = {15732},
  year    = {1999},
  doi     = {10.1103/PhysRevB.60.15732}}

@article{Moskalets2002,
  author  = {M. Moskalets and M. B\"uttiker},
  title   = {Floquet scattering theory of quantum pumps},
  journal = {Phys. Rev. B},
  volume  = {66},
  pages   = {205320},
  year    = {2002},
  doi     = {10.1103/PhysRevB.66.205320}}

@article{Pantazopoulos2019,
  author  = {P. A. Pantazopoulos and N. Stefanou},
  title   = {Layered optomagnonic structures: Time {F}loquet scattering-matrix approach},
  journal = {Phys. Rev. B},
  volume  = {99},
  pages   = {144415},
  year    = {2019},
  doi     = {10.1103/PhysRevB.99.144415}}

@article{Kiorpelidis24,
  title = {Transient amplification in stable Floquet media},
  author = {Kiorpelidis, Ioannis and Diakonos, Fotios K. and Theocharis, Georgios and Pagneux, Vincent},
  journal = {Phys. Rev. B},
  volume = {110},
  issue = {13},
  pages = {134315},
  numpages = {10},
  year = {2024},
  month = {Oct},
  publisher = {American Physical Society},
  doi = {10.1103/PhysRevB.110.134315},
  url = {https://link.aps.org/doi/10.1103/PhysRevB.110.134315}
}

@misc{garg2026BIC,
      title={Bound states in the continuum in multilayered time-varying metasurfaces}, 
      author={Puneet Garg and Michael Plum and Carsten Rockstuhl},
      year={2026},
      eprint={2607.06469},
      archivePrefix={arXiv},
      primaryClass={physics.optics}
}

@article{Suwunnarat,
  title = {Dynamically modulated perfect absorbers},
  author = {Suwunnarat, Suwun and Halpern, Dashiell and Li, Huanan and Shapiro, Boris and Kottos, Tsampikos},
  journal = {Phys. Rev. A},
  volume = {99},
  issue = {1},
  pages = {013834},
  numpages = {7},
  year = {2019},
  month = {Jan},
  publisher = {American Physical Society},
  doi = {10.1103/PhysRevA.99.013834},
  url = {https://link.aps.org/doi/10.1103/PhysRevA.99.013834}
}

@article{Li18,
  title = {Floquet-Network Theory of Nonreciprocal Transport},
  author = {Li, Huanan and Kottos, Tsampikos and Shapiro, Boris},
  journal = {Phys. Rev. Appl.},
  volume = {9},
  issue = {4},
  pages = {044031},
  numpages = {13},
  year = {2018},
  month = {Apr},
  publisher = {American Physical Society},
  doi = {10.1103/PhysRevApplied.9.044031},
  url = {https://link.aps.org/doi/10.1103/PhysRevApplied.9.044031}
}

@article{Li19,
  title = {Design Algorithms of Driving-Induced Nonreciprocal Components},
  author = {Li, Huanan and Kottos, Tsampikos},
  journal = {Phys. Rev. Appl.},
  volume = {11},
  issue = {3},
  pages = {034017},
  numpages = {9},
  year = {2019},
  month = {Mar},
  publisher = {American Physical Society},
  doi = {10.1103/PhysRevApplied.11.034017},
  url = {https://link.aps.org/doi/10.1103/PhysRevApplied.11.034017}
}

@article{Brizard,
  title = {Local {M}anley-{R}owe Relations for Noneikonal Wave Fields},
  author = {Brizard, A. J. and Kaufman, A. N.},
  journal = {Phys. Rev. Lett.},
  volume = {74},
  issue = {23},
  pages = {4567--4570},
  numpages = {0},
  year = {1995},
  month = {Jun},
  publisher = {American Physical Society},
  doi = {10.1103/PhysRevLett.74.4567},
}

@article{Leonhardt,
    doi = {10.1088/0034-4885/66/7/203},
    url = {https://dx.doi.org/10.1088/0034-4885/66/7/203},
    year = {2003},
    month = {jun},
    publisher = {},
    volume = {66},
    number = {7},
    pages = {1207},
    author = {Ulf Leonhardt},
    title = {Quantum physics of simple optical instruments},
    journal = {Rep. Prog. Phys.},
}

@article{Bellotti,
	author = {Bellotti, U. and Bornatici, M. and Engelmann, F.},
	date = {1997/05/01},
	doi = {10.1007/BF02897900},
	id = {Bellotti1997},
	isbn = {1826-9850},
	journal = {Riv. Nuovo Cimento},
	number = {5},
	pages = {1--67},
	title = {Radiative energy transfer in anisotropic, spatially dispersive, weakly inhomogeneous and dissipative media with embedded sources},
	volume = {20},
	year = {1997},
}

@article{Koutserimpas20,
	title        = {Electromagnetic Fields in a Time-Varying Medium: Exceptional Points and Operator Symmetries},
	author       = {Koutserimpas, Theodoros T. and Fleury, Romain},
	year         = 2020,
	journal      = {IEEE Trans. Antennas. Propag.},
	volume       = 68,
	number       = 9,
	pages        = {6717--6724},
	doi          = {10.1109/TAP.2020.2996822}
}

@article{Koutserimpas183,
	title        = {Parametric amplification and bidirectional invisibility in $\mathcal{PT}$-symmetric time-{F}loquet systems},
	author       = {Koutserimpas, Theodoros T. and Al\`u, Andrea and Fleury, Romain},
	year         = 2018,
	month        = {Jan},
	journal      = {Phys. Rev. A},
	publisher    = {American Physical Society},
	volume       = 97,
	pages        = {013839},
	doi          = {10.1103/PhysRevA.97.013839},
	url          = {https://link.aps.org/doi/10.1103/PhysRevA.97.013839},
	issue        = 1,
	numpages     = 8
}

\end{document}